\documentclass[aps,proof,showpacs, twocolumn, floatfix]{revtex4}%
\usepackage{amsfonts}%
\usepackage{amsmath}%
\usepackage{amssymb}%
\usepackage{graphicx}

\begin{document}

\title{Density Functional Theory of Inhomogeneous Liquids: III. Liquid-Vapor
Nucleation}
\author{James F. Lutsko}
\affiliation{Center for Nonlinear Phenomena and Complex Systems CP 231, Universit\'{e}
Libre de Bruxelles, Blvd. du Triomphe, 1050 Brussels, Belgium}
\email{jlutsko@ulb.ac.be}
%\pacs{61.20.Gy,05.70.Np,68.03.-g}

\begin{abstract}
The process of nucleation of vapor bubbles from a superheated liquid and of
liquid droplets from a supersaturated vapor is investigated using the
Modified-Core van der Waals model Density Functional Theory (Lutsko,
JCP 128, 184711 (2008)). A novel approach is developed whereby
nucleation is viewed as a transition from a metastable state to a
stable state via the minimum free energy path which is
identified using the nudged elastic-band method for exploring free
energy surfaces.  This allows for the unbiased calculation of the
properties of sub- and super-critical clusters, as well as of the
critical cluster. For Lennard-Jones fluids, the results compare well to simulation and support
the view that even at high supersaturation nucleation proceeds
smoothly rather than via spinodal-like instabilities as has been
suggested recently.
\end{abstract}

\date{\today }
\maketitle

\section{Introduction}

The liquid-vapor phase transition is of fundamental importance in many areas
of physics, chemistry and engineering. For this reason, the process of
homogeneous nucleation of bubbles in a superheated fluid and of droplets in
a supersaturated vapor has a long history going all the way back to the
fundamental paper of van der Waals\cite{VDW1,VDW2,RowlinsonWidom}. Recent years have seen tremendous
progress in the development of simulation methods that circumvent the time
and size limitations of simulations to  allow for the study of rare events,
such as homogeneous nucleation\cite{Reiss3,frenkel_gas_liquid_nucleation,OhZeng1,FrenkelSmit}. In parallel, various approaches
to a theoretical description of nucleation have been formulated with
varying degrees of success\cite{Oxtoby_Evans,talanquer:5190, Corti_PRL}. However, the theoretical description
of nucleation is difficult because it is fundamentally a
nonequilibrium process: the initial and final states are metastable
and stable, respectively, but the transition between them involves a sequence of
unstable configurations making the use of equilibrium statistical
mechanics difficult.

Density Functional Theory (DFT) is the modern realization of van der
Waals' approach to inhomogeneous fluids. The fundamental quantity in
DFT is the ensemble-averaged local density, $\rho(\mathbf{r})$ in the
presence of an external field, $\phi(\mathbf{r})$ at fixed
temperature, $T$ and chemical potential, $\mu$. It can be proven
that there is a one to one relationship between applied fields and the
resulting density profiles. Furthermore, it can also be shown that
there exists a functional, $F[n]$ such that the quantity $\Omega[n] \equiv F[n] + \int n(\mathbf{r})
\left( \phi(\mathbf{r}) - \mu \right) d\mathbf{r}$ is minimized when $n(\mathbf{r})$ is the
equilibrium density profile, $\rho(\mathbf{r})$ corresponding to the
applied field\cite{MerminDFT, HansenMcdonald}. Then, $\Omega[\rho]$ is the grand potential for the
given field, temperature and chemical potential. Since the theory only
gives meaning to density functions which minimize $\Omega[n] $, the
question is how to extract information from this formalism in the case
of nucleation in the absence of an external field (or, perhaps, at
constant gravitational field). 

Intuitively, one imagines that if the functional $\Omega[\rho]$ gives the free energy for any density profile 
$\rho(\mathbf{r})$ in the presence of the appropriate stabilizing field, then, in the absence of such a field, 
it should give information about the energetic cost of moving from a stable state to the state $\rho(\mathbf{r})$. 
For closed systems, this intuition has been formalized in recent years in the development known as dynamical density functional 
theory\cite{Evans_1979_AdvPhys, TarazonaDDFT1, EvansDDFT} and, more generally, it underlies phase-field models
for both open and closed systems (see, e.g. ref.\cite{TpperGrant_EPL_2008} for a recent example). For all of these models, when fluctuations are included, 
detailed balance implies that the probability of observing a density profile $\rho(\mathbf{r})$ is proportional to $e^{-\beta L[\rho]/\sigma^2}$ 
where  $L[\rho] = F[\rho]$ for closed systems and $L[\rho] = \Omega[\rho]$ for open systems and where $\sigma$ characterizes the strength of fluctuations. 
One then expects that an approximation to the most likely path from one state to another in an open system will be 
the  shortest path involving the smallest free energy barriers which is to say the Minimal Free Energy Path (MFEP) as defined by $\Omega[\rho]$.
This of course underlies the classical theory of nucleation and is the view adopted in the present work. 

The simplest method to approximate the MFEP is to define a class of
parametrized profiles, $\rho(\mathbf{r}) = \rho(\mathbf{r}; R,w,
...)$ where $R$ is the radius of the bubble or droplet, $w$ is its
width and the notation indicates that there may be additional
parameters as well. Then, the functional $F[\rho]$ becomes simply a
function of the parameters $R$, $w$, ... and one can define a path by
minimizing it for each fixed value of $R$ with respect to the
remaining parameters. The problem, of course, is that one restricts
the space of possible density profiles and may exclude the most likely
profile. A second possibility is to introduce some auxiliary
constraint that defines what is meant by a bubble (or droplet) of a
given size. For example, such a constraint for a bubble could be
$\int \left( \rho_{l}-\rho(\mathbf{r}) \right) d\mathbf{r} = \Delta N$
which  defines a profile to be of size $\Delta N$ if the total number
of atoms is $\Delta N$ less than in the uniform fluid. For reasons
described elsewhere\cite{LutskoEPL2008}, this particular example of  a constraint is
not really useful but more realistic constraints have been
constructed\cite{talanquer:5190,Corti_PRL,CortiJCP}. Note however the philosophy of this approach: a path
through density-function space is defined by the parameter $\Delta N$
(or whatever parameters enter the constraint). This is not so
different in spirit from the use of parametrized profiles as both
methods effectively reduce the dimensionality of the problem. However,
in the case of gas bubbles, it has been shown using both numerical
calculations and analytic models that the constraint approach can fail
 to yield
a stable profile for a fixed value of $\Delta N$ (or whatever
parameters are used) even though the
underlying free energy landscape is well behaved\cite{LutskoEPL2008}. Hence, while it
is certainly more general than the use of parametrized profiles, the
constraint method is not a robust approach to the determination of the
MFEP.

In fact, the search for minimum energy paths over some energy
landscape is a problem that occurs in many applications such as the
determination of chemical reaction paths from ab initio calculations
and the determination of transitions between cluster
structures\cite{Wales}. In recent years, several techniques have been
developed for determining the MFEP. Here, one such method - the Nudged
Elastic Band (NEB)\cite{NEB,NEB-CI} - is applied to the problem of nucleation. This
allows for a completely unbiased and robust determination of the MFEP
between two metastable states. 

In this paper, the details of droplet and bubble nucleation are calculated for the
Lennard-Jones fluid. Because
the structure of the liquid-vapor interface is a balance between bulk
free energy and surface tension, it is important that a useful theory
be able to describe both quantities with quantitative accuracy. The
calculations presented here therefore  use
the Modified-Core van der Waals (MC-VDW) model DFT\cite{lutsko_jcp_2008_2} which gives a
quantitatively accurate description of the planar liquid-vapor
interface, of the structure of the LJ fluid near a hard wall and of
the LJ fluid confined to a slit pore\cite{lutsko_jcp_2008_2}. In the next section, the
model is described and the details of the application of the NEB
method to the problem of bubble nucleation is outlined. The results of the calculations are
presented in  Section
III. There evidence is given of the robustness of the method and direct, quantitative comparison is made to the results of computer simulation.
The MFEP for nucleation of droplets from a supersaturated vapor and of bubbles from a supercooled liquid are described. The last Section gives a summary of the results obtained and their implication concerning recent claims of non-standard pathways in liquid-vapor nucleation.

\section{Theory}
\subsection{Density Functional Theory}
The model free energy functional used in my calculations is the
Modified-Core van der Waals model\cite{lutsko_jcp_2008_2} which is
written as a sum of three contributions,
\begin{equation}
F[\rho] = F_{id}[\rho] + F_{hs}[\rho] + F_{core}[\rho] + F_{tail}[\rho].
\end{equation}
The first contribution is the ideal gas term which is given by
\begin{equation}
F_{id}[\rho] = \int \left( \rho(\mathbf{r}) \log\left( \rho(\mathbf{r}) \right) -\rho(\mathbf{r}) \right) d\mathbf{r}.
\end{equation}
Next is a hard-sphere contribution, $F_{hs}[\rho]$,
for which the ``White Bear'' Fundamental Measure Theory (FMT) model was
used\cite{WhiteBear, tarazona_2002_1} along with the Barker-Hendersen
hard-sphere diameter\cite{BarkerHend,HansenMcdonald}. The third contribution,
the ``core correction'' $F_{core}[\rho]$,
is similar to a FMT model but is constructed so that the total free
energy functional reproduces a given equation of state in the bulk
phase as well as certain other conditions concerning the direct
correlation function in the bulk fluid\cite{lutsko_jcp_2008_2}. The
final term is a mean-field treatment of the long-range attraction,
\begin{equation}
F_{tail}[\rho] = \int \Theta(r_{12}-d) \rho(\mathbf{r}_{1})
\rho(\mathbf{r}_{2})v(r_{12}) d\mathbf{r}_{1} d\mathbf{r}_{2},
\end{equation}
where $\Theta(x)$ is the step function, $d$ is the Barker-Henderson
hard-sphere diameter and $v(r)$ is the pair potential. In most of the
calculations described below, the potential is a truncated and shifted
Lennard-Jones (LJ) interaction,
\begin{equation}
v(r)\,=\,\left\{\begin{array}
{r@{\quad:\quad}l}
v_{LJ}(r) -\,v_{LJ}(r_c)& r\,<\,r_c\\
0 & r\geq r_c
\end{array}\right.
\end{equation}
where $v_{LJ}(r) =
4\epsilon\left(\left(\frac{\sigma}{r}\right)^{12}-\left(\frac{\sigma}{r}\right)^{6}\right)$
is the untruncated LJ potential. (In one case, comparison is made to simulations performed with an unshifted potential and in that case the calculations were performed with a truncated, but unshifted potential.) The DFT model requires as input the bulk equation of state. Since the object of the
calculations was to model the LJ system as accurately as possible, the
empirical equation of state of Johnson, Zollweg and Gubbins
\cite{JZG} was used. 

\subsection{Determining the MFEP}
The NEB method is a chain-of-states description of the MFEP. A path in density space is described by a collection of profiles, $\{\rho^{a}(\mathbf{r})\}_{a=0}^M$. To be concrete, consider the problem of the nucleation of bubbles in a superheated liquid. Then, the initial state is the uniform liquid, $\rho^{(0)}(\mathbf{r}) = \rho_l$ where $\rho_l$ is the bulk liquid density determined by the temperature and chemical potential. The subsequent images are initialized to a guess of the MFEP: for example, a sequence of hyperbolic tangents with increasing radii. Of course, if one simply tries to minimize the total energy of the path, $\sum_{a=1}^{M-1} \Omega[\rho^{(a)}]$ then nothing is learned since the images will eventually converge to one of the two attractors, the uniform liquid or the uniform vapor, depending on whether their initial values are smaller than, or larger than, the critical cluster. The starting point of the NEB method is the addition of fictitious elastic forces between the images in order to force them to remain evenly spaced along the MFEP. The following discussion is divided into two parts: first the general implementation of the NEB is described after which the specialization to spherical symmetry is given.

The basic element needed to implement the method is a definition of a scalar product in density space. The natural choice is, for two real functions $f(\mathbf{r})$ and $g(\mathbf{r})$,  
\begin{equation}
f * g  =  \int f(\mathbf{r})g(\mathbf{r}) d\mathbf{r} 
\end{equation}
Note that the limits of the integral are not specified: for most purposes here, the product will only be evaluated for functions having finite support so that some large volume will suffice. Using this definition, the distance between two profiles, $\rho^{(i)}(\mathbf{r})$ and $\rho^{(j)}(\mathbf{r})$ is defined as
\begin{equation}
\left| \rho^{(i)} - \rho^{(j)} \right|^2 = \left( \rho^{(i)} - \rho^{(j)} \right) * \left( \rho^{(i)} - \rho^{(j)} \right)
\end{equation}
which is the usual $L^2$-norm. The idea behind the NEB is to minimize the free energy functionals, $\Omega[\rho^{i}]$, in the manifold orthogonal to the current estimate of the MFEP and to move them along the estimated MFEP using fictitious elastic forces to maintain an even spacing of the images. To this end, the critical element is the estimation of the tangent to the MFEP at each density image for which the algorithm of ref. \cite{NEB} was used. This involves the neighboring images and their free energies. For example, if $\Omega[\rho^{(i-1)}] < \Omega[\rho^{(i)}] < \Omega[\rho^{(i+1)}]$, then the tangent to the image $\rho^{(i)}$, called $t^{(i)}$, is 
\begin{equation}
t^{(i)}(\mathbf{r}) = \rho^{(i+1)}(\mathbf{r}) - \rho^{(i)}(\mathbf{r}) 
\end{equation}
and the normalized tangent, $\hat{t}^{(i)}(\mathbf{r}) = t^{(i)}(\mathbf{r})/\left(t^{(i)}*t^{(i)}\right)$. If the inequalities are reversed, the tangent is in the direction $\rho^{(i)}-\rho^{(i-1)}$. For non-monotonic neighbors, the heuristic is given in ref.\cite{NEB}. The NEB method then consists of finding a configuration that gives zero NEB-force. Let the ``force'' due to the actual free-energy surface be $\mathcal{F}^{(i)}(\mathbf{r}) = -\frac{\partial \beta \Omega[\rho^{(i)}]}{\partial \rho(\mathbf{r})}$. Then the NEB method consists of solving
\begin{equation} \label{NEB}
0 = \mathcal{F}^{\perp(i)}(\mathbf{r}) + k\hat{t}^{(i)}(\mathbf{r})\left( \left| \rho^{(i+1)}-\rho^{(i)} \right| - \left| \rho^{(i)}-\rho^{(i-1)} \right| \right)
\end{equation}
where $\mathcal{F}^{\perp(i)}(\mathbf{r}) = \mathcal{F}^{(i)}(\mathbf{r}) - \hat{t}^{(i)}(\mathbf{r})\left(\hat{t}^{(i)} * \mathcal{F}^{(i)}\right)$ is the component of the thermodynamic force orthogonal to the tangent vector and $k$ is the spring constant. 

The specialization to spherical geometry is made by noting that if the density is spherically symmetric, $\rho(\mathbf{r}) = \rho(r)$, then the corresponding thermodynamic forces, $\mathcal{F}(\mathbf{r})$ and $\mathcal{F}_S(r) \equiv  -\frac{\partial \beta \Omega[\rho^{(i)}]}{\partial \rho(r)}$, are related by the functional chain rule,
\begin{align}
\mathcal{F}_S(r)  & =  - \int \frac{\partial \beta \Omega[\rho^{(i)}]}{\partial \rho(\mathbf{r'})} \frac{\partial \rho(\mathbf{r'})}{\partial \rho(r)} d\mathbf{r'} \\
 & =  - \int \frac{\partial \beta \Omega[\rho^{(i)}]}{\partial \rho(\mathbf{r'})} \delta(r'-r) d\mathbf{r'} \notag \\
 & =  4 \pi r^2 \mathcal{F}(\mathbf{r}) \notag 
\end{align}
It therefore follows that
\begin{widetext}
\begin{align}
t^{(i)}(r)  = &\rho^{(i+1)}(r) - \rho^{(i)}(r) \\
0  = & \mathcal{F}_S^{\perp(i)}(r) + 4 \pi r^{2} k\hat{t}^{(i)}(r)\left( \left| \rho^{(i+1)}-\rho^{(i)} \right| - \left| \rho^{(i)}-\rho^{(i-1)} \right| \right) \notag \\
\mathcal{F}_S^{\perp(i)}(r)  = & \mathcal{F}_S^{(i)}(r) - 4 \pi r^2 \hat{t}^{(i)}(r) \left(\hat{t}^{(i)} * \mathcal{F}^{(i)}\right) \notag \\
\hat{t}^{(i)} * \mathcal{F}^{(i)} = & \int \hat{t}^{(i)}(r)  \mathcal{F}^{(i)}(\mathbf{r}) d \mathbf{r} \notag \\
 = & \int \hat{t}^{(i)}(r)  \mathcal{F}_S^{(i)}(r) dr \notag
\end{align}
\end{widetext}
and so forth. Note that this differs somewhat from the more heuristic scheme described in ref. \cite{LutskoEPL2008}. The present approach, starting from the general case and specializing to spherical symmetry, is more systematic and yields lower free energies than did the earlier implementation. 

A final refinement is the use of a so-called ``climbing image''\cite{NEB-CI}. This is an image, say image $\rho^{(c)}$ for which the sign of the component of $\mathcal{F}^{(c)}$ along the path is reversed, instead of eliminating this component, thus causing it to climb towards the local maximum. For this image, no spring forces are applied so that its behavior is governed by
\begin{equation}
 -\mathcal{F}_S^{\perp (c)}(r) + \left( \mathcal{F}_S^{(c)}(r) - \mathcal{F}_S^{\perp (c)}(r)\right) = 0.
\end{equation}
(Note, however, that the images on either side of the climbing image still feel spring forces connecting them to the climbing image: there is no analogy of Newton's third law in this case.) The climbing image proves very effective in determining the saddle point, which is to say in this problem, the critical cluster. 

So far, nothing definite has been said about the endpoint of the chain of states. The initial point is the uniform liquid, but the end cannot be taken to be the uniform gas because the distance between the uniform gas and any finite bubble is infinite. In an initial implementation of this method\cite{LutskoEPL2008}, the end point was taken to be a sigmoidal profile with fixed, large radius and with the width adjusted to minimize the free energy. The idea was that the profiles of large bubbles would be basically sigmoidal and any error introduced by the approximation would be insignificant.  However, this is somewhat unsatisfying as it potentially biases the chain of states away from the MFEP. In the present work, a more elegant approach was taken. Note that in a part of the chain of states where the free energy is monotonically decreasing, as it is for clusters larger than the critical cluster,  the tangent vector for the image $i$ is determined by the images $i-1$ and $i$. The only role played by image $i+1$ is in the calculation of the spring force on image $i$. However, if this image were excluded completely, the spring between $i$ and $i-1$ would cause these images to converge. To avoid this, it is sufficient to replace the term $\left| \rho^{(i+1)}-\rho^{(i)} \right|$ in Eq.(\ref{NEB}) by any convenient constant. Then, there is no harm in terminating the chain even though the last image is neither fixed, nor a minimum of the free energy. This gives a completely unbiased approach to the determination of the MFEP.

\subsection{Relation to previous approaches}
As far as I am aware, the present use of the NEB method for exploring nucleation in the context of DFT is novel.  However, the problem has been studied using other approaches. The properties of the critical cluster are in principle accessible since it is an extremum of the free energy and several authors have calculated its properties, albeit using less sophisticated DFT models than that employed here\cite{Oxtoby_Evans,OxtobyZengJCP1991,CortiJCP}. Since the critical cluster is a saddle point, rather than a minimum, in the free energy surface, it is still difficult to isolate. One would like to make an initial guess as to its structure and then to solve the Euler-Lagrange equation which inevitably involves an iterative refinement of the initial guess until stability is reached. However, because the critical cluster is a saddle point, and because the only stable minima are the uniform liquid and vapor, one finds that an initial guess that is too small evolves towards smaller and smaller clusters until the uniform state is reached, while one that is too large evolves towards larger and larger clusters until the uniform state of the other phase is reached. Oxtoby and Evans therefore used this behavior to bracket the critical cluster with successive initial guesses and extracted the properties of the critical cluster at an intermediate point in the iterative calculations\cite{OxtobyZengJCP1991}. This technique gives correct results but is obviously messy and the NEB with the climbing image can be seen as an improvement since it gives the critical cluster in one run.

The properties of non-critical clusters have been studied by Lee, Telo da Gama and Gubbins\cite{Lee} and by Talanquer and Oxtoby\cite{talanquer:5190}. In the study of Lee et al., finite-sized volumes where used. By varying the size of the volume and/or the density of material within the volume, clusters of different sizes become stable - presumably because the system wants to phase-separate. The problem in this case is that it is not clear what the analogies can be drawn between these clusters and those forming in open systems. Talanquer and Oxtoby used a method inspired by this approach and developed by Reiss et al\cite{Reiss1, Reiss2, Reiss3}. In the language used in this paper, the Talanquer-Oxtoby method consists of minimizing the grand potential, $\Omega[n]$, subject to the constraint that the total number of atoms in a volume $v$ be fixed at a given number, $i$,
\begin{equation}
i = \int_v \rho(\mathbf{r}) d\mathbf{r} .
\end{equation}
(The relation between this formulation and that of ref.\cite{talanquer:5190} is given in the Appendix.) Talanquer and Oxtoby furthermore adjust the density outside the volume so that the density profile is continuous. The idea here is, in some sense, to reduce the number of degrees of freedom required to explore density function space to just two, the parameters $i$ and $v$, thus reducing the difficulty of the problem of finding the MFEP. However, as shown in the Appendix, this amounts to changing the chemical potential so as to stabilize a cluster with the prescribed number of atoms. In other words, the clusters so obtained are just the critical clusters at different chemical potentials. These would not appear to be physically equivalent to clusters of different sizes at fixed chemical potentials. Uline and Corti have used a similar method to explore bubble nucleation but without the adjustment of the density outside the clusters\cite{Corti_PRL}. This circumvents the objection raised above, but the method has other problems: the generation of discontinuous density profiles and the appearance of spurious instabilities\cite{LutskoEPL2008}. Thus, the constraint method, while physically appealing, is not a reliable approach for the exploration of density-function space. 

\subsection{Parametrized profiles : Classical nucleation theory}
A somewhat simpler alternative to minimizing the free energy with respect to a density function is to use parametrized density profiles of the form $n(\mathbf{r};\Gamma)$ where $\Gamma$ represents one or more parameters. An example is the hyperbolic tangent profile,$n(\mathbf{r};A,z_0) =  \rho_l + \frac{1}{2}(\rho_v-\rho_l)(1+\tanh(-A(z-z_0)))$ used to described planer interfaces. For a spherical geometry, a closely related form,
\begin{equation} \label{hyper}
\rho\left( r\right)=\rho _{l}+\left( \rho _{v} -\rho _{l}\right) \frac{1+br}{%
1+\left( br\right) ^{2}}\frac{1-\tanh \left( A(r-R)\right) }{1-\tanh \left(
-A_{}R\right) }
\end{equation}
has the advantage of having zero derivative at the origin and of showing the expected behavior  $n(r) \sim exp(-ar)/r$ at large $r$\cite{GuntonProtein}. This profile will be used as an initial guess for the NEB calculation.

An even simpler parametrization is of particular importance. Using $n(r,R) = \rho_l\Theta(R-r) + \rho_v\Theta(r-R)$, where the step function $\Theta(x)$ is one for $x>0$ and zero otherwise, in conjunction with a simple van der Waals free energy functional gives
\begin{align}
\Delta \beta \Omega(R;\mu, T) & = \frac{4 \pi}{3}R^3(\beta P(\rho_v;T) - \beta P(\rho_l;T)) \\
& + 4 \pi R^2 \gamma_{coex} \left( \frac{\rho_l-\rho_v}{\rho_{l,coex}-\rho_{v,coex}} \right)^2 \nonumber
\end{align} where $P(\rho)$ is the bulk pressure at density $\rho$, $\gamma_{c,coex}$ is the surface tension at coexistence and $\rho_{l,coex}$ and $\rho_{v,coex}$ are the liquid and vapor densities at coexistence\cite{LutskoEPL2008}. This is just a slight generalization of the CNT model for the free energy of a bubble. (The classical model results from setting $\rho_{l,coex}-\rho_{v,coex} = \rho_l-\rho_v$ as will be done henceforth.) This expression for the excess free energy has minima at $R=0$ and $R \rightarrow \infty$ corresponding to the uniform vapor and liquid, respectively and a maximum at the critical radius,
\begin{equation}
R_c = \frac{2 \gamma_{coex}}{\beta P(\rho_v;T) - \beta P(\rho_l;T)}
\end{equation} and 
\begin{equation}
\Delta \beta \Omega(R_c;\mu, T) = \frac{16 \pi}{3} \frac{\gamma_{coex}^3}{(\beta P(\rho_v;T) - \beta P(\rho_l;T))^2}
\end{equation}

\subsection{Computational details}
In the calculations presented below, the radial variable was discretized into 40 points per hard-sphere diameter. The profiles were discretized out to a maximum radius of 20 hard-sphere diameters giving 800 points in total. The contributions to the free energy coming from the density profiles at larger distances were taken into account by assuming the density to be constant and equal to the appropriate uniform (liquid or vapor) density at larger distances. As alluded to above, the results were very insensitive to the value of the spring constant, which was, in most cases, fixed at $1$ in LJ units. For calculations very close to the spinodal, where the energy barriers are very small, a smaller spring constant sometimes proved to give better convergence. 

The modified Euler-Lagrange equation was solved using the fast inertial relaxation engine\cite{FIRE}. In this algorithm, a fictitious time variable is introduced and each component of each image is treated as a dynamical variable moving in response to the ``forces''. The algorithm involves a quench so that the system relaxes toward a state of zero force. This algorithm appears to be one of the most efficient ways to implement the NEB method\cite{NEB-FIRE}. In the present case, it was found to be necessary to treat the images as being dynamically independent: that is to say, each image had its own time variable, time step and quenching variables. The parameters governing the quenches were the same as those given in ref.\cite{FIRE} and the quench was deemed to be converged when the root-mean-squared force was less than $5 \times 10^{-5}$ in LJ units when only the energies were of interest. The convergence of the profiles, particularly near the core, $r=0$, are slow because of the $r^2$ weighting of the forces so the algorithm was run until the rms force was less than $1 \times 10^{-5}$ when the profiles were desired. 

\section{Results}
\subsection{General aspects droplet nucleation}
\begin{figure*}[htpb]
\begin{center}
\resizebox{12cm}{!}{
{\includegraphics[height = 12cm, angle=-90]{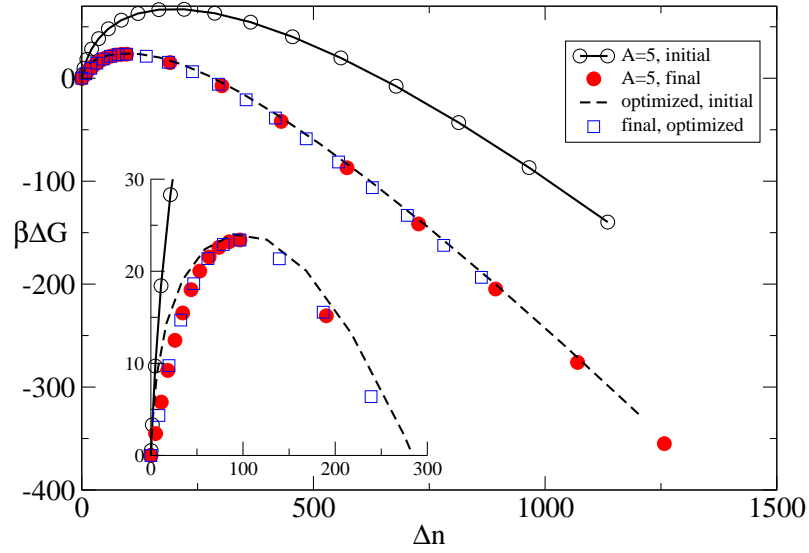}}}
\end{center}
\caption{(Color on line) Initial and final free energy barriers for
  two different initial conditions: $A=5$ and optimized hypertangent.}
\label{fig1}
\end{figure*}

In this section, the nucleation of liquid droplets from a supersaturated vapor will be examined. Before turning to specific examples which will mainly involve a comparison to existing simulation data, it is interesting to consider some general properties of the present method. Two important properties of the critical cluster are its size and excess free energy. The size of a cluster may be unambiguously defined as the number of atoms in the system relative to the number in the metastable state,
\begin{equation} \label{excess}
\Delta n = \int \left( \rho(\mathbf{r}) - \rho_{\infty} \right) d\mathbf{r}
\end{equation}
where in the case of droplets (bubbles) $\rho_{\infty}$ is the density of the bulk vapor (liquid)  at the specified chemical potential and temperature. 
Figure \ref{fig1} shows the size and free energy barrier of the critical droplets calculated for the LJ potential with cutoff $r_c^* =4$ at temperature $T= 0.8$ and excess chemical potential $\mu-\mu_{coex} = 0.15\mu_{coex}$, corresponding to a supersaturation of $S_P\equiv P_v/P_{coex}=2.27$. (Note that with this cutoff, the critical point is calculated to occur at $T_c=1.25$.) The calculations was performed using 20 images which were initialized as modified hyperbolic tangents, Eq.(\ref{hyper}), with increasing radii. The figure shows the results of calculations in which the initial guess of the width of the interfaces was very narrow ($A=5$ in Eq.(\ref{hyper})) which is far from the optimized-sigmoidal value ($A \sim 1$) as well as the result using initial guesses near the optimized-sigmoidal value. It is apparent that, even though the initial paths are very different, the final results are indistinguishable in shape. Note that one of the final curves is longer than another because the choice of which image is the ``climbing image'', that locates the critical droplet, was based on the initial energies and these differed in the two runs. Since the climbing image always ends up at the maximum of the free energy, the number of images before and after the maximum can differ if the initial curves are sufficiently different. This figure also shows that the final free energy barrier is indistinguishable from the initial optimized-sigmoidal profile for large clusters which just confirms that large clusters are indeed sigmoidal in shape. However, as shown in the inset, the converged barrier is definitely lower than the optimized sigmoidal barrier. 

\begin{figure*}[htpb]
\begin{center}
\resizebox{12cm}{!}{
{\includegraphics[height = 12cm, angle=-90]{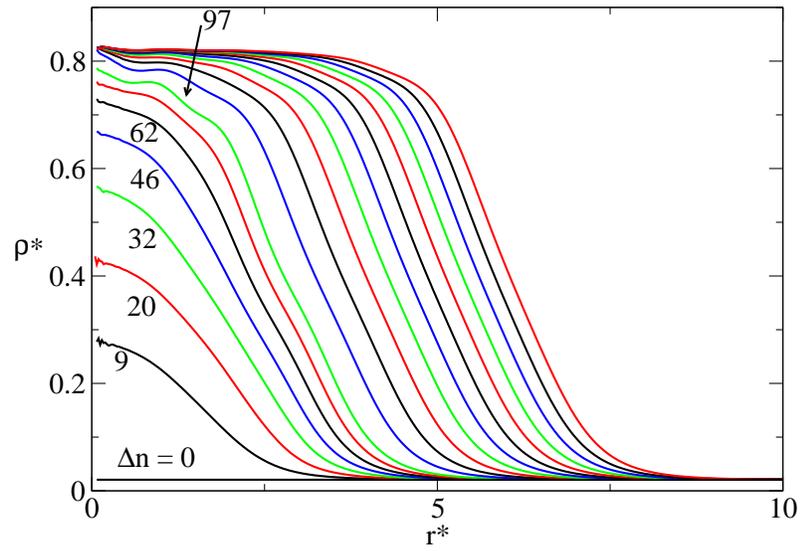}}}
\end{center}
\caption{(Color on line) Profiles of subcritical and supercritical droplets for $T^*=0.8$ and $S_P=2.27$. The system starts in a uniform state with density $\rho^*(r) \equiv \rho(r)\sigma^3 = 0.0207$ corresponding to an excess number of atoms $\Delta n = 0$. The various curves represent various points along the MFEP. The first few curves are labeled with the value of $\Delta n$ and the critical cluster is indicated with an arrow.}
\label{fig2}
\end{figure*}
\begin{figure*}[htpb]
\begin{center}
\resizebox{12cm}{!}{
{\includegraphics[height = 12cm, angle=-90]{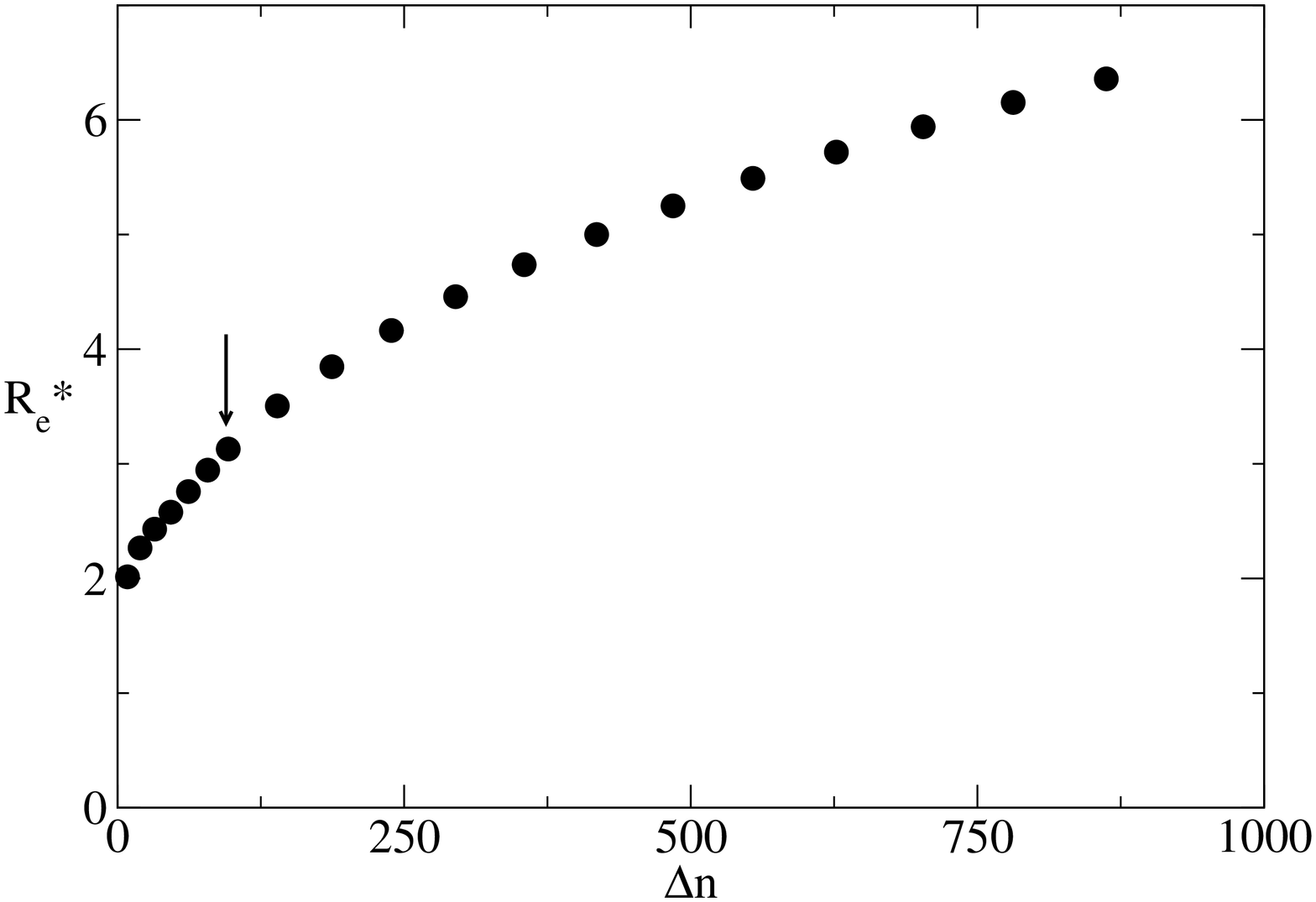}}}
\end{center}
\caption{The equimolar radius as a function of the excess number of
  atoms in a droplet for $T^*=0.8$ and $S_P=2.27$. The numbers give $\Delta n$ for the indicated clusters. The cluster marked with the arrow is the critical cluster.}
\label{fig3}
\end{figure*}

Since the calculations sample the entire nucleation pathway, a picture of the development of droplets is given. Figure \ref{fig2} shows the sequence of droplets along the MFEP for the same conditions as above. The figure indicates that the droplets form via a gradual increase in the density near the core. This is in contrast to the picture in CNT where small droplets are assumed to have the density of the bulk liquid over a small volume. In fact, as the figure shows, even very small droplets appear to have finite volume. This can be quantified by calculating the equimolar radius, defined as the radius of a sphere with uniform density $\rho(0)$ that has the same excess number of atoms as the actual density profile,
\begin{equation}
\frac{4 \pi}{3} R_e^3 \left(\rho(0) - \rho_{\infty} \right) = \Delta n .
\end{equation} 
This  is shown in Fig. \ref{fig3} and the results clearly support the contention that the radius of the droplets does not go to zero.

\subsection{Comparison to simulation}
There have been many simulation studies of the formation of liquid clusters in a supersaturated vapor. Simulations of stable clusters are possible using constant particle number, volume and temperature (NVT): see e.g. \cite{thompson:530,Lee,Mareschal,OhZeng2}. Beginning with an unstable density, the system will spontaneously phase separate into one or more liquid droplets surrounded by vapor. However, this does not correspond to nucleation of droplets in the laboratory which normally occurs as constant pressure or at constant volume but with volumes so large that the vapor pressure remains nearly constant. In contrast, the NVT simulations have mostly been performed on small systems so as to result in the formation of a single droplet and the size of the droplet and density of the surrounding vapor is then a function of the size of the simulation cell. One exception is the work of Oh and Zeng\cite{OhZeng1, OhZeng2} which was done using relatively large systems. Comparison to their work will be made below.

Perhaps the cleanest simulations, from the standpoint of a comparison to theory, have been those of ten Wolde and Frenkel(tWF) carried out in the constant number, pressure and temperature - or NPT -ensemble\cite{tWF}. By means of umbrella sampling, they were able to determine the properties of unstable clusters of all sizes, at least for one value of the supersaturation. Since their method gives a faithful sampling of the NPT ensemble, direct comparison of theory to simulation is relatively straight forward, with some caveats. For example, the DFT calculations are performed
in the grand ensemble: physically, these are expected to give the same
results as in both cases, droplets are nucleated within a background
vapor of essentially constant density. More formally, it is easy to
see that the free energy \emph{differences} should be the same in both
ensembles\cite{Oxtoby_Evans}. 
\begin{figure*}[htpb]
\begin{center}
\resizebox{12cm}{!}{
{\includegraphics[height = 12cm, angle=-90]{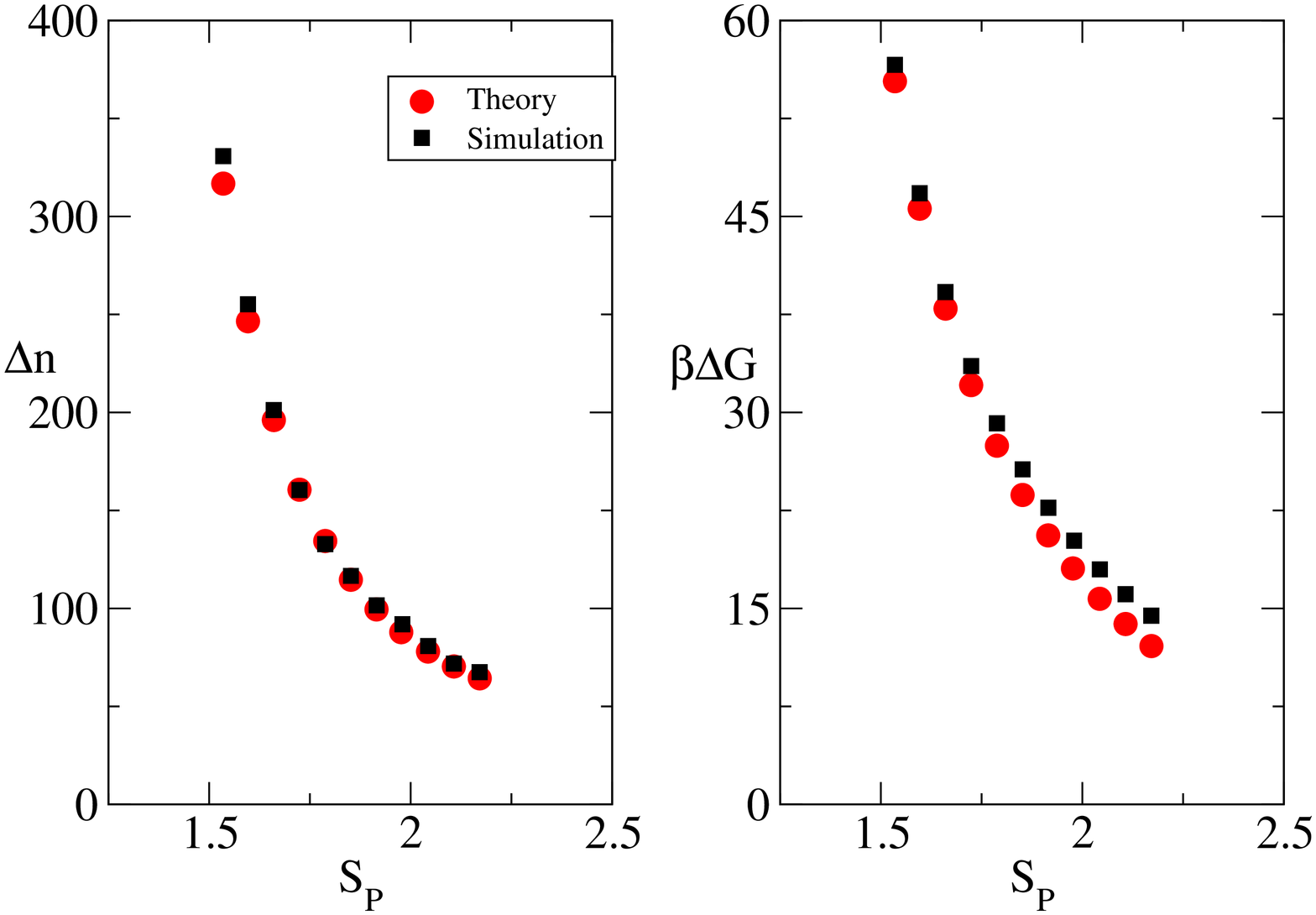}}}
\end{center}
\caption{(Color on line) Comparison of theory (circles) and simulation
  results  of ten Wolde and Frenkel\cite{tWF} (squares) for (a) the excess number of atoms in the critical
  cluster and (b) the free energy barrier height as functions of the
  supersaturation.} 
\label{fig4}
\end{figure*}
tWF used a truncated
and shifted potential with cutoff $r_c=2.5\sigma$ and all simulations
were performed at a reduced temperature of $T^*\equiv k_BT/\epsilon =0.741$.  In the following,
the reduced density is $\rho^* \equiv \rho \sigma^3$ and the reduced length is $r^* \equiv r/\sigma$. 

As stated
above, the MC-VDW model was solved using the empirical JZG equation of
state. However, the JZG EOS was developed for an infinite ranged
potential. The usual means to take account of a finite-ranged
potential is to introduce mean-field corrections\cite{JZG}. For relatively large
cutoffs, this gives an accurate account of the thermodynamics but for
cutoffs as short as that used by tWF, the corrected equation of state
is not very accurate. In particular, it gives a critical temperature
of $T^*_c=1.0366$ whereas simulation gives $T^*_c=1.085$. This
difference of about $4\%$ is important as the surface tension goes to
zero at the critical temperature and is thus very sensitive to its
value. Fortunately, it has been shown that the effect of this
inaccuracy can be almost completely eliminated by invoking the law of
corresponding states and comparing theory and simulation at equal
values of $T/T_c$\cite{Grosfils}. This is the strategy used here so
that the theory is evaluated at $T/T_c = 0.741/1.085 = 0.683$ or,
given the theoretical value of the critical temperature,
$T^*=0.708$. 
\begin{figure*}[htpb]
\begin{center}
\resizebox{12cm}{!}{
{\includegraphics[height = 10cm, angle=-90]{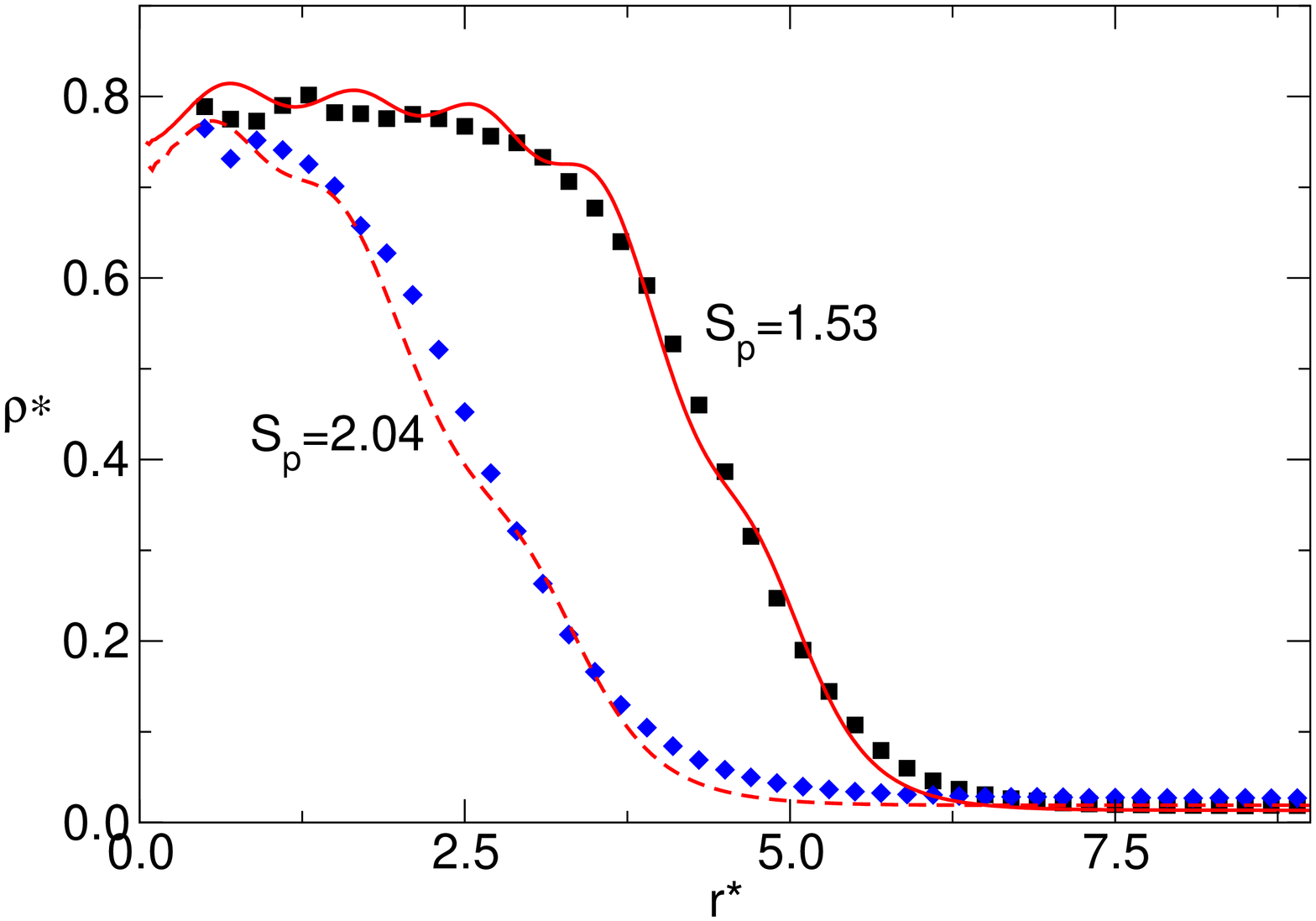}}}
\end{center}
\caption{(Color on line) Comparison of theory (lines) and the simulation
  results of ten Wolde and Frenkel\cite{tWF} (symbols) for the structure of the critical cluster for two different
values of the supersaturation, $S_p$.}
\label{fig5}
\end{figure*}

Figure \ref{fig4} shows the excess number of atoms in the critical
cluster and the free energy barrier as
functions of the supersaturation, $S_P=P_v/P_{coex}$ where $P_v$ is the
pressure in the bulk vapor and $P_{coex}$ is the pressure at
coexistence. The agreement is very good even at quite high
supersaturations and correspondingly small clusters and is seen to be far better
than the predictions of CNT. The agreement for small clusters is
particularly important as the clusters are so small, with a radius less than $2\sigma$, 
that virtually all atoms in the cluster are part of the
interface: in other words, the system is extremely
inhomogeneous. Figure \ref{fig5} shows the theoretical prediction for
the structure of the critical cluster at $S_P=1.535$ compared to the
simulation results. In both cases, the structure near $r=0$ is poorly resolved: in the simulations because there are few atoms and poor statistics, and in the calculations because the $r^2$ weighting in the integrals means that this region has very little effect on the free energy. Thus, ten Wolde and Frenkel only report the density profile for $r^* > 1$ in ref.\cite{tWF} although more data is shown in the figure here. The theory is clearly very accurate in giving the
correct overall size and shape of the density profile. It does show
some structure near the core that is absent from the simulation but
this is because the theory does not take into account the smearing of
the interface expected to result from both center of mass
motion\cite{talanquer:5190} and capillary waves\cite{Katsov}. 
\begin{figure*}[htpb]
\begin{center}
\resizebox{12cm}{!}{
{\includegraphics[height = 12cm, angle=-90]{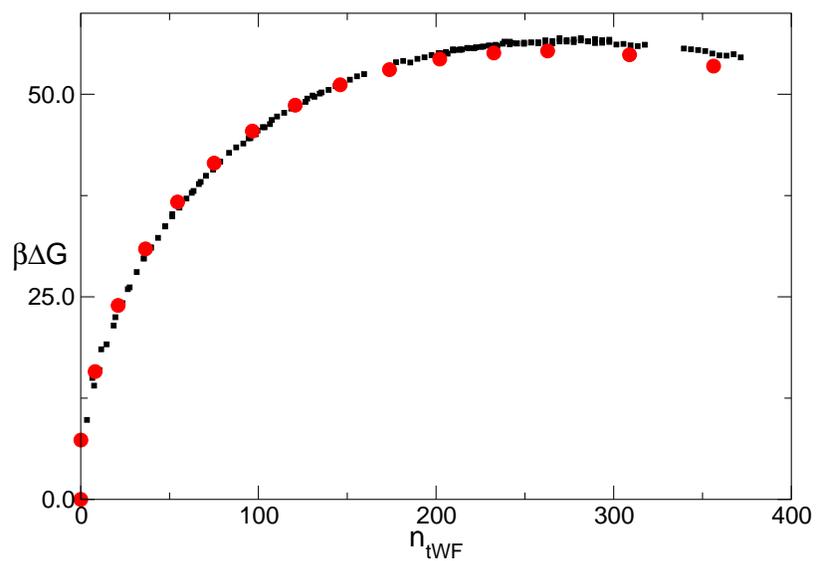}}}
\end{center}
\caption{(Color on line) The nucleation barrier at supersaturation
  $S_P=1.535$ as a function of cluster size. The squares are from
  simulation  of ten Wolde and Frenkel\cite{tWF} and the circles are from the theory.}
\label{fig6}
\end{figure*}

All of the properties compared so far are restricted to the
critical nucleus. However, one of the most impressive aspects of the
work of tWF is that they were able to determine the free energy
barrier as a function of cluster size for one particular value of
supersaturation ($S=1.535$). This allows for a check of the novel
approach used here to determine the MFEP as well as its relevance to
the nucleation problem. There is, however, one subtlety in making this
comparison. tWF characterized the barrier as a function of
\emph{cluster size} rather than the excess number of atoms in a
cluster. Their definition of whether or not an atom was in the liquid
or vapor was based on the local density: an atom was classified as
part of a liquid cluster if it had at least 4 neighbors within a
distance of $q=1.5\sigma$. There is no practical, exact way to
translate this into a criteria that can be evaluated theoretically so
the following heuristic procedure was used. The number of neighbors of
within a distance $q$ of an atom  in the uniform bulk system at
density $\rho$ is $n = 4 \pi \int_0^q \rho g(r;\rho) r^2 dr$ where
$g(r,\rho)$ is the pair distribution function(PDF). I have used this
expression, together with the usual first order
Weeks-Chandler-Anderson perturbative approximation for the
PDF\cite{WCA1,WCA2, WCA3, HansenMcdonald}, to determine that $n \ge 4$
occurs for $\rho^* > 0.32$. Hence, in the theory, all regions
with density satisfying this inequality were classified as liquid. For
the critical cluster, where theory gives and excess number of atoms of
$315$ compared to $330$ reported by tWF, this procedure gives a
theoretical cluster size of $265$ compared to $285$ found in the
simulations. It would appear that this is a sensible way to calculate
``cluster size'' in the theory. 

Figure \ref{fig6} shows a comparison of the nucleation barrier as a
function of cluster size as determined from simulation and theory. The
two are in remarkable agreement with the theoretical values about $1.5 k_BT$
smaller than those observed in the simulation at the barrier's maximum. This is consistent with the fact
that the theory determines the MFEP whereas the simulations report a
thermal average which will also involve nearby, higher energy,
states. This agreement gives strong empirical validation of the
present theoretical approach.

\begin{figure*}[htpb]
\begin{center}
\resizebox{12cm}{!}{
{\includegraphics[height = 12cm, angle=-90]{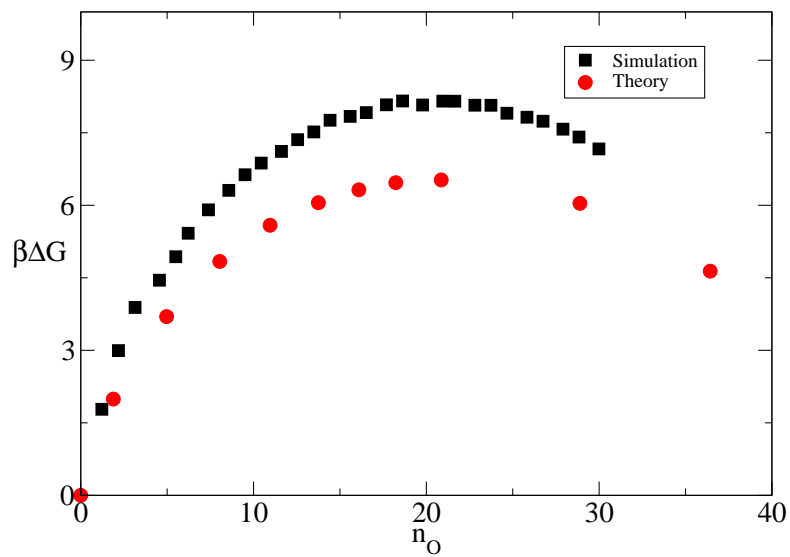}}}
\end{center}
\caption{(Color on line) The nucleation barrier as determined from simulation by Oh and Zeng\cite{OhZeng1}, squares, and from the theory, circles. The definition of the cluster size, $n_O$, is given in the text.}
\label{fig7}
\end{figure*}

Oh and Zeng have performed a set of Monte Carlo simulations of large systems at high supersaturation where the object was to generate an equilibrium distribution of clusters from which the free energy barrier can be extracted\cite{OhZeng1,OhZeng2}. Their work is complementary to that of ten Wolde and Frenkel in
 that it focuses on  higher supersaturations, and consequently much smaller clusters. However, direct comparison of Oh and Zeng's results
to theory is complicated by their definition of the supersaturation which involves a sum over the entire population of clusters. In one case, however, they
do report the pressure of the ambient vapor as $P\sigma^3/(k_BT) = 0.01166$ at $T^*=0.67$\cite{OhZeng2}. Since the potential cutoff used in the simulations was $r_c=4.5\sigma$ (with no shift), it can be assumed that the JZG equation of state with a mean-field correction for the cutoff is essentially exact and no corresponding-states adjustments are necessary. In this case, the main uncertainty comes from the definition of the cluster size, denoted $n_O$. Oh and Zeng define two atoms to be in the same cluster if they are within $1.5\sigma$ of one another. The theoretical cluster size was therefore calculated as in the case of tWF except the limiting density was chosen to be that at which there is less than one neighbor in a sphere of radius $1.5\sigma$. Figure \ref{fig7} shows the free energy barrier determined from simulation compared to that calculated here. In particular, the barrier height is calculated to be $6.5k_BT$ compared to the value from simulation of $8.2k_BT$ while the peak occurs at almost the same place, $n_O \sim 20$. Although the relative error in the barrier height is greater than in the case of the tWF data, the comparison is actually remarkably similar in that the theoretical barrier is lower than the observed barrier by almost exactly the same amount found in the comparison to tWF while the cluster size is virtually identical to that found in the simulations, again as found in the comparison to tWF. Given the very small cluster size, the uncertainty in the thermodynamic state and in the estimate of the cluster size, and the fact that the simulations involve many clusters while the calculations are for an isolated cluster, the agreement is probably as good as could be expected.
\begin{figure*}[htpb]
\begin{center}
\resizebox{12cm}{!}{
{\includegraphics[height = 12cm, angle=-90]{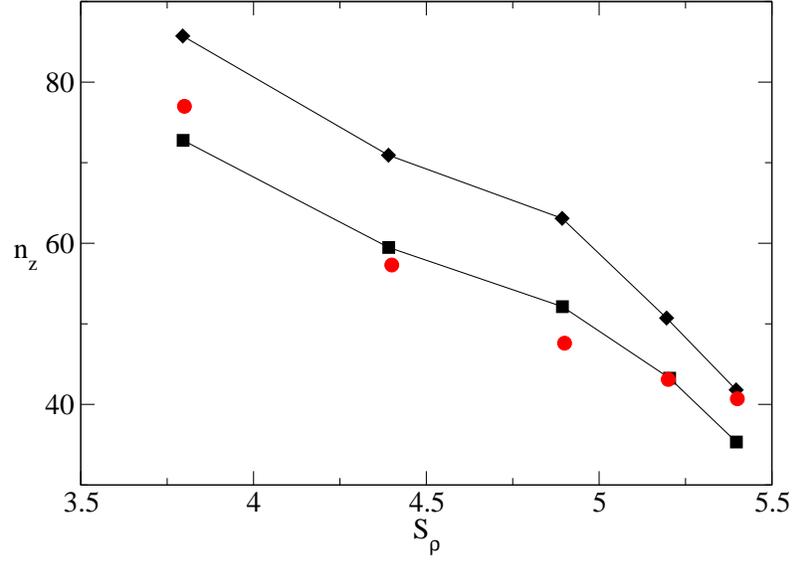}}}
\end{center}
\caption{(Color on line) The size of the critical nucleus as a function of supersaturation. The squares and diamonds are the lower and upper bound on the size of the nucleus as determined by Zhukovitskii\cite{Zhuk} (the lines are a guide to the eye) and the circles are the results of the calculations described in the text.}
\label{fig8}
\end{figure*}

The comparison to the results of Zhukovitskii\cite{Zhuk} is more problematic for two reasons. First, there was no fixed cutoff in the simulations: the potential was essentially infinite-ranged but the volume is finite so that the number of neighbors a given atom interacted with depended on how close that atom was to the (spherical) cell boundary. The radius of the cell was $R_Z=8\sigma$ so the effective cutoff is in the range $0 \le r_c \le 8\sigma$. Zhukovitskii reports the density of the vapor at coexistence as being $\rho_{v,coex} = 0.00243$ so I have adjusted the cutoff to approximately match this, which happens at $r_c \sim 5\sigma$. Clearly, this is only an approximation and probably represents the greatest source of error. Next, Zhukovitskii reports the supersaturation in terms of the variable $S_{\rho} = \rho_v / \rho_{v,coex}$ however, this was only estimated assuming the vapor is an ideal gas: apparently, the actual vapor pressure was typically about $4\%$ lower than this\cite{Zhuk}, so in the calculations, it was assumed that $\rho_v = 0.95 S_{\rho} \rho_{v,coex}$. With such a large cutoff, the error in the equation of state should be minimal so there was no need for the kinds of adjustment required for the tWF cutoff.  Given these caveats, Fig. \ref{fig8} shows the size of the critical cluster as a function of supersaturation as reported by Zhukovitskii compared to the calculations. Given the uncertainty in the cutoff and in the actual density of the vapor, not to mention the fact that the simulation does not correspond precisely to any standard ensemble, the agreement is satisfactory.

\subsection{Nucleation of Bubbles}
\begin{figure*}[htpb]
\begin{center}
\resizebox{12cm}{!}{
{\includegraphics[height = 12cm, angle=-90]{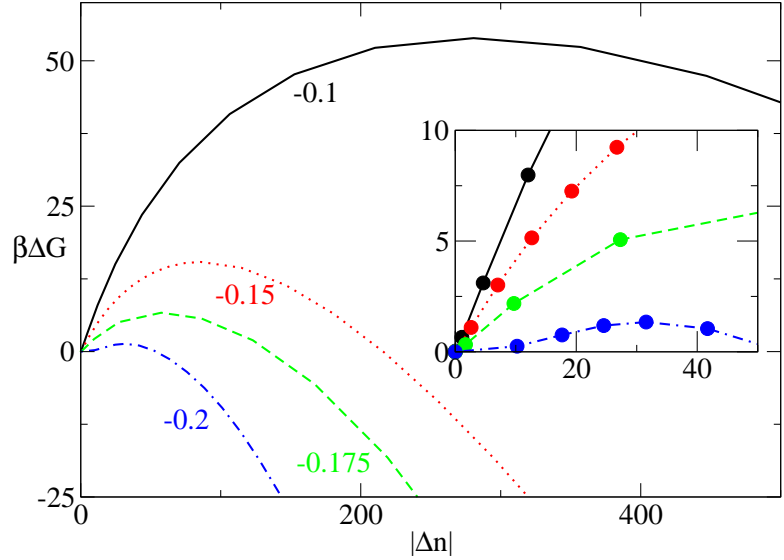}}}
\end{center}
\caption{(Color on line) The free energy barrier for bubble nucleation as a function of bubble size at several different values of supersaturation as calculated from the theory. The curves are labeled with their supersaturation, $S_{\mu}.$. The inset shows an expansion of the small $\Delta n$ region.}
\label{fig9}
\end{figure*}

\begin{figure*}[htpb]
\begin{center}
\resizebox{12cm}{!}{
{\includegraphics[height = 12cm, angle=-90]{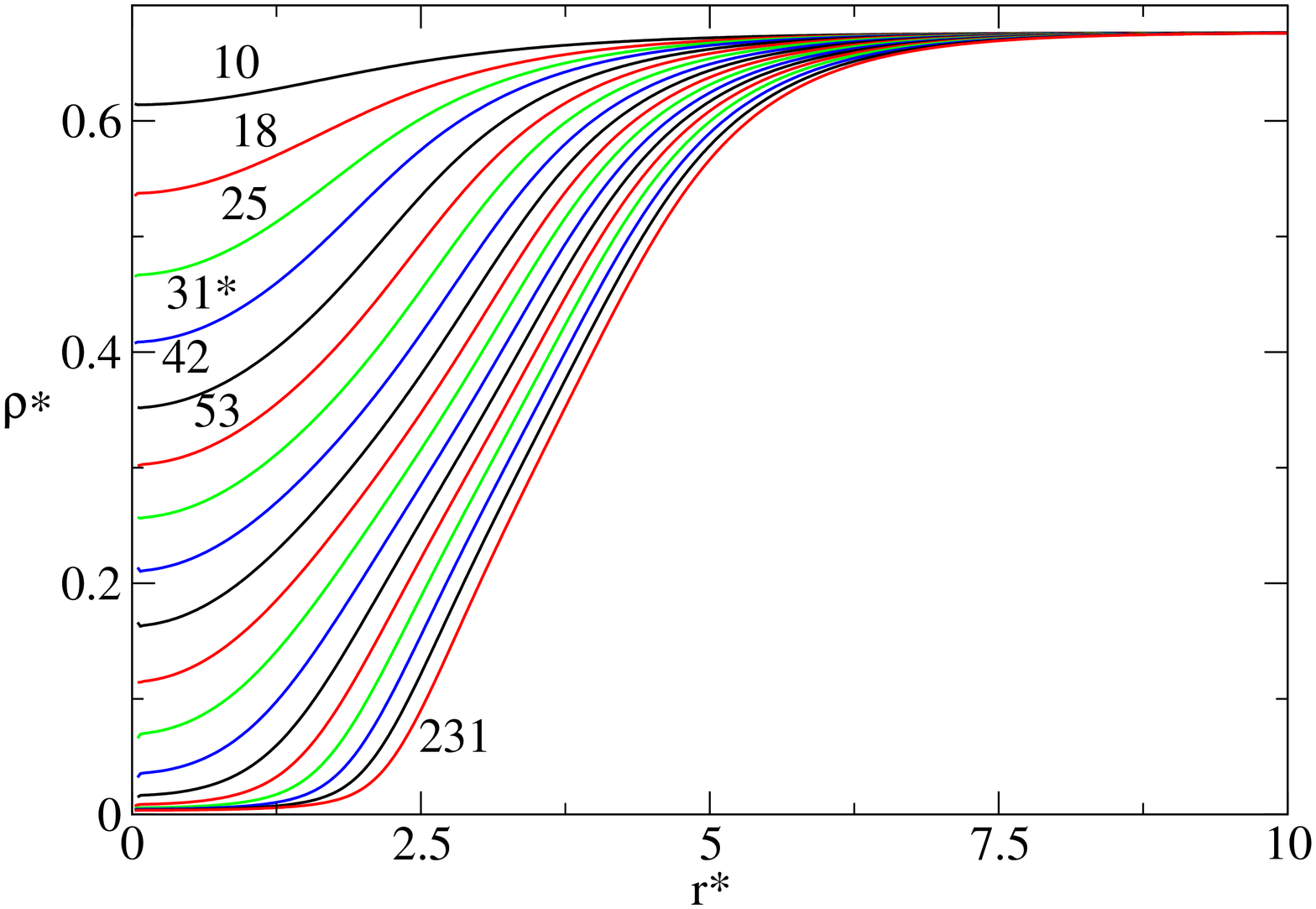}}}
\end{center}
\caption{(Color on line) A sequence of bubble profiles along the MFEP for supersaturation $S_{\mu}=-0.20$. The size, in terms of atomic deficit or $-\Delta n$, is given for several profiles and the critical profile is marked with an asterisk.}
\label{fig10}
\end{figure*}

The same methods can be used to study the nucleation of bubbles in a superheated liquid. Figure \ref{fig9} shows the height of the nucleation barrier as a function of bubble size for different supersaturations for $T^*=0.8$ and a cutoff of $r_c^*=4.0$. Note that for bubbles, $\Delta n$ is negative and the supersaturation is given here as $S_{\mu}\equiv \frac{\mu-\mu_{coex}}{\mu_{coex}}$. . In contrast to recent claims that there is an activated instability in bubble nucleation\cite{Corti_PRL}, the results indicate continuous paths for a wide range of supersaturations. It is possible that at high absolute value of supersaturation the free energy is concave near $\Delta n = 0$ but there is no sign of a non-classical instability, even when the free energy barrier is less than $2k_BT$.  Figure \ref{fig10} shows a sequence of bubbles for $S_{\mu}=-0.15$. As in the case of droplet nucleation, the width is apparently always greater than zero with the bubble nucleating via an initial, gradual lowering of the central density followed by a slow broadening into a typical sigmoidal shape. Note that the density at the center of the critical cluster shown in Fig. \ref{fig10} is more than half the liquid density and, hence, much greater than the density of the gas phase being nucleated. This represents a large difference from the assumption of CNT that the critical cluster has the density of the bulk vapor inside the bubble. 

\section{Discussion}
The nucleation of a stable phase from a metastable phase is best
understood as a transition between two local minima in the free energy
surface. As such, it is conceptually the same as any problem that
involves the crossing of a free energy barrier and, in particular,
bears strong similarities to the description of  chemical
reactions. It is therefore not surprising that the methods used
to determine reaction pathways can be usefully applied to the problem
of nucleation. 

There are in fact many methods used to study reaction pathways
including eigenvalue-following\cite{Wales, WalesPRB}, the string method\cite{string}, the determination
of the maximum likelihood path\cite{MaxLikelihoodPath} among others, including the NEB method used
here. The NEB method may not be the optimal method for nucleation, but
it has the advantage of being very simple to implement and of being
robust.

There has been much discussion recently concerning the possibility of
non-classical mechanisms of liquid-vapor and of vapor-liquid
nucleation. In particular, Bhimalapuram, Chakrabarty and Bagchi have
observed a spinodal-like breakdown at high supersaturations in Ising
model simulations of condensation\cite{bhimalapuram:206104} although the significance
of their observation has recently been
questioned\cite{maibaum:019601}. Similarly, Uline and Corti have
claimed a similar type of behavior for boiling based on a combination of
simulations and DFT calculations\cite{Corti_PRL}. A similar
observation concerning crystallization has been made by
Parinello\cite{Parrinello}. What all of these studies have in common is
the observation of nonstandard elements in the free energy surface:
either an additional local minimum\cite{bhimalapuram:206104} or some
type of discontinuity\cite{Corti_PRL}. The present work supports none
of these observations. In all cases, even at very high
supersaturation, the free energy surface appears to be well-defined
and to possess no non-classical minima. For the nucleation of droplets from a vapor, 
these results are consistent with the simulations of ten Wolde and Frenkel\cite{tWF} and of Oh and Zeng\cite{OhZeng1}, 
neither of which saw evidence of non-classical features in the free energy barrier. In the case of the nucleation
of bubbles in a superheated liquid, the calculations may indicate that at very high supersaturations, such that the free
energy barrier is less than $2 k_BT$, the free energy barrier is not
a convex function of its size, but it is not clear that this is of any
significance. Furthermore, there is evidence, based on simple
analytically tractable models, that the discontinuities observed by
Uline and Corti are an artifact of the constraint method used to
explore the free energy surface\cite{LutskoEPL2008}. Thus, while the present results
cannot decisively settle the question one way or another, they do tend
to support the classical picture of liquid-vapor nucleation as a
simple matter of barrier crossing.

\begin{acknowledgments}
I am grateful to Pieter ten Wolde and Daan Frenkel for supplying their simulation data. This work was supported in part by the European Space Agency under contract
number ESA AO-2004-070.
\end{acknowledgments}

\appendix{}

\section{The Constraint Method}
The constraint method described by Talanquer and Oxtoby consists of demanding that the number of atoms within the volume $v$ be fixed at some value $i$. Furthermore, the density outside the volume is prescribed to be a fixed value, $\rho_{\infty}$. Minimizing the grand potential under these constraints can be formulated using Lagrange multipliers so that one minimizes the functional
\begin{eqnarray}
F[n] - \mu \int n(\mathbf{r}) d\mathbf{r} -\alpha \left( \int_{r<R} n(\mathbf{r}) d\mathbf{r} -i \right) \\
- \int_{r>R} \gamma(\mathbf{r})\left( n(\mathbf{r}) -\rho_{\infty}\right) d\mathbf{r} \notag
\end{eqnarray} 
with respect to $n(\mathbf{r})$, $\alpha$, and $\gamma(\mathbf{r})$. For $r < R$ this gives
\begin{align}
0 & = \frac{\delta F[n]}{\delta n} - \mu  -\alpha  \\
0 & = \int_{r<R} n(\mathbf{r}) d\mathbf{r} -i \notag 
\end{align}
so that it is clear that the effect of the Lagrange multiplier is simply to shift the chemical potential. The resulting density profile is just that of the system at the shifted chemical potential. To make contact with the work of Talanquer and Oxtoby, it is necessary to separate the functional derivative occurring in this expression into its ideal gas and excess contributions by writing
\begin{equation}
\frac{\delta F[n]}{\delta n} = k_BT \log n(\mathbf{r}) + \mu_{ex}(\mathbf{r})
\end{equation}
This allows the first of the variational equations to be rearranged to give
\begin{equation}
n(\mathbf{r}) = \exp \left( \beta \mu  + \beta \alpha- \beta \mu_{ex}(\mathbf{r}) \right)
\end{equation}
Substitution into the second variational equation gives an expression for the Lagrange Multiplier
\begin{equation}
0  = \exp(\beta \alpha)\int_{r<R} \exp \left( \beta \mu  - \beta \mu_{ex}(\mathbf{r}) \right) d\mathbf{r} -i \notag
\end{equation} 
or
\begin{equation}
\beta \alpha + \beta \mu = \log(i) - \log\left( \int_{r<R} \exp \left(- \beta \mu_{ex}(\mathbf{r}) \right) d\mathbf{r}\right)
\end{equation} 
Thus, the equation for the density, for the case $r<R$, becomes
\begin{equation}
\frac{\delta \beta F[n]}{\delta n} = \log(i) - \log\left( \int_{r<R} \exp \left(- \beta \mu_{ex}(\mathbf{r}) \right) d\mathbf{r}\right)
\end{equation}
which is, essentially, Eq. 17 of ref.(\cite{talanquer:5190}).

The variational equations for $R < r$ are
\begin{align}
0 & = \frac{\delta F[n]}{\delta n} - \mu  - \gamma(\mathbf{r})  \\
0 & = n(\mathbf{r}) -\rho_{\infty} \notag
\end{align}
for very large $r$, far away from the interface, it will be the case that 
\begin{equation}
V^{-1}\left( \frac{\delta F[n]}{\delta n} \right)_{n=\rho_{\infty}} \sim \frac{\partial f(\rho_{\infty})}{\partial \rho_{\infty}}
\end{equation}
so that in this region, $\mu + \gamma(\mathbf{r}) \rightarrow \frac{\partial f(\rho_{\infty})}{\partial \rho_{\infty}} = \mu(\rho_{\infty})$, where the last term means the chemical potential corresponding to the density $\rho_{\infty}$. In particular, if $R$ is sufficiently large that the cluster interface is located at much smaller values of the radius, as appears to be the case in the work of Talanquer and Oxtoby, then a continuous density profile will basically require that $\alpha + \mu = \gamma(\mathbf{r}) + \mu = \mu(\rho_{\infty})$ so that the procedure simple amounts to an overall shift of the chemical potential.

\bigskip 
\bibliographystyle{apsrev}
\bibliography{nucleation}

\end{document}